\documentclass[a4paper,11pt]{article}
\usepackage{jheppub}
\usepackage{amsmath}
\usepackage{amsfonts}
\usepackage{amssymb}
\usepackage{graphicx}
\usepackage{wrapfig}
\usepackage{braket}
\usepackage{hyperref}
\usepackage{amsthm,verbatim,cancel}
\usepackage{color}
\usepackage{caption}
\usepackage{subcaption}
\newcommand{\bea}{\begin{eqnarray}}
\newcommand{\eea}{\end{eqnarray}}
\newcommand{\nn}{\nonumber\\}

\ifx\color@rgb\@undefined\else
  \definecolor{IK}{rgb}{0, 0.51, 1}
\fi

\begin{document}

\title{Entanglement Hamiltonians and entropy in 1+1D chiral fermion systems}

\author{Israel Klich, Diana Vaman and Gabriel Wong}
 \affiliation
    {%
    Department of Physics,
    University of Virginia,
    Charlottesville, VA 22904, USA
    }%
\abstract{

In past work \cite{Klich:2015ina} we introduced a method which allows for exact computations of entanglement Hamiltonians. The method relies on computing the resolvent for the projected (on the entangling region) Green's function using a solution to the Riemann-Hilbert problem combined with finite rank perturbation theory. Here we analyze in detail several examples involving excited states of chiral fermions (Dirac and Majorana) on a spatial circle. We compute the exact entanglement Hamiltonians and an exact formula for the change in entanglement entropy due to the introduction of a particle above the Dirac sea. For Dirac fermions, we give the first-order temperature correction to the entanglement entropy in the case of a multiple interval entangling region.}

\maketitle

\section{Introduction}

Entanglement Hamiltonians are effective Hamiltonians that arise when we wish to view a part of a system as a thermal state, even when the rest of the system does not correspond to a proper thermal bath. For closed systems at a finite temperature, these effective Hamiltonians equal, of course, to the ordinary Hamiltonians. However, in general, entanglement Hamiltonians are quite different from typical local Hamiltonians appearing in physics. In particular, entanglement Hamiltonians contain information about the entanglement between the subsystem and its complements, including the Schmidt spectrum and Schmidt eigenvalues, and as a consequence entanglement entropy, its variants and other quantities.
The interest in the nature of entanglement Hamiltonians and in their computation is therefore steadily growing.  
The problem is of considerable interest as it is fundamental to our understanding of the interplay between locality and correlations in quantum fields. As such it is being addressed in a range of studies from high energy physics to condensed matter and quantum information \cite{casini2009reduced,peschel2011relation,lauchli2012entanglement,qi2012general,hermanns2014entanglement,poilblanc2014entanglement,Lashkari:2015dia,rispler2015long,cardy2016entanglement}. Indeed, entanglement Hamiltonians are a new object to explore in our quest to understand the quantum structure of field theories and many-body states. 

The entanglement Hamiltonians are defined through the reduced density matrix as follows. The reduced density matrix $\rho_{V}$  on a spatial region $V$ is defined to reproduce expectation values of all operators $O_{V}$ localized inside $V$ via the relation 
\bea
\rm Tr_{V}(\rho_{V} O_V )  = \langle O_{V} \rangle,
\eea 
where  $ \langle O_{V} \rangle= Tr \rho~O_V $ is the quantum mechanical and statistical expectation value of $O_V$ in the full system. Often the state of the full system will be taken as a pure state, but we shall also consider low temperature corrections.

The entanglement Hamiltonian $\mathcal{H}_V$ is then given as the effective Hamiltonian inside $V$ that satisfies
\bea
 \rho_{V}=\frac{e^{-\mathcal{H}_V}}{Z_V},
\eea
where $Z_V$ is a normalization factor such that $\text{Tr} \rho_V=1$. 
Or, simply, $\mathcal{H}_V=-\log (Z_V\rho_{V})$.

There have been precious few successes in the exact calculation of entanglement Hamiltonians due to the difficulty of the problem. However, progress has been made in several cases. Most notably, conformal field theories with a simple spherical entangling region have an entanglement Hamiltonian that can be directly expressed in terms of the free space Hamiltonian density \cite{bisognano1976duality,myers,Wong:2013fk}. Other cases have also been explored. In particular, the modular Hamiltonian for chiral fermions on the plane has been computed in \cite{casini2009reduced}. In \cite{Klich:2015ina} a method for computing entanglement Hamiltonians for fermions in 1+1d has been proposed by relating it to a Riemann-Hilbert problem, and used to compute the entanglement Hamiltonians for 1+1d chiral fermions on a circle with and without zero modes.

Here we proceed to use the method for several new applications. In particular, we find  exact entanglement Hamiltonians for simple excited states and thermal states of free fermions in a low temperature expansion.  Previously, the entanglement entropy for free Dirac fermions at finite size and temperature was computed via the replica trick and bosonization in \cite{Azeyanagi:2007bj} and \cite{herzog2013entanglement}.  This result for the entanglement entropy on a single interval was generalized to conformal field theories at finite temperature in \cite{Cardy:2014jwa} in a low temperature expansion. Meanwhile a replica trick calculation of the single interval entanglement Hamiltonian was developed in \cite{Lashkari:2015dia}. Our method involves a direct calculation of the fermion entanglement Hamiltonian and provides an alternative to the replica trick calculations. As such we are able to tackle a multiple interval entangling region with the same set of tools as for the single interval case.  

The paper is structured as follows. In section \ref{subsec:DiracExcited}, we compute the entanglement Hamiltonian for a basic excited state of chiral Dirac fermions where an additional particle has been introduced above the Dirac sea. We use the same method to compute the change in entropy relative to the entanglement entropy of the Dirac sea. Here we derive an exact formula for the entropy change, equation \eqref{delta sv}, and compare it to a numerical evaluation using a system of particles hopping on a ring. In section \ref{subsec:MajoranaExcited} we repeat the problem for the case of Majorana fermions. The treatment here involves using exactly solvable rank-2 perturbation theory, and we show how to extend the original method to accommodate higher rank perturbations. In section \ref{sec:DiracLowTemp} we find the low-temperature correction to the entanglement entropy. In particular, we find that the first low-temperature correction to the entanglement entropy formula for multiple intervals remains formally the same as for the single interval, with $L$ now being interpreted as the total length of the entangling region. We present additional details of the method in the appendices. In Appendix \ref{apres}  we give a pedagogical introduction to some of the algebraic structure behind the Riemann-Hilbert approach. In Appendix \ref{sec:AppendixMatrixElements} we employ this algebraic structure to compute various matrix elements that we use throughout the paper.

%%%%%%%%%%%%%%%%%%%%%%%%%%%%%%%%%%%%%%%%%%%%%%%%%%%%%%%%%%%%%%%%
%%%%%%%%%%%%%%%%%%%%%%%%%%%%%%%%%%%%%%%%%%%%%%%%%%%%%%%%%%%%%%%%%%%%%%%
\section{Entanglement Hamiltonians and entanglement entropy for excited states of chiral fermions}\label{sec:DiracExcited}
%%%%%%%%%%%%%%%%%%%%%%%%%%%%%%%%%%%%%%%%%%%%%%%%%%%%%%%%%%%%%%%%%%%%%%%
In the following sections we give a straightforward application and extension of the method introduced by us in \cite{Klich:2015ina}, and derive the entanglement Hamiltonian for  certain excited state of chiral fermions.
%%%%%%%%%%%%%%%%%%%%%%%%%%%%%%%%%%%%%%%%%%%%%%%%%%%%%%%%%%%%%
\subsection{Rank one perturbations: chiral Dirac Fermions}\label{subsec:DiracExcited}

Here, we consider a chiral Dirac fermion in 1+1 dimensions, with the  spatial dimension a circle of radius $R$. We are free to impose generic boundary conditions $\Psi(x+2\pi R)=\Psi(x) e^{2\pi \alpha i}$, with $0\leq\alpha <1$. Correspondingly the fermions have the mode expansion
\bea
&&\Psi(x,t)=\frac{1}{\sqrt{2\pi R}}\sum_{r\in Z+\alpha} a_r e^{-i\tfrac rR(t-x)}=
e^{-i\tfrac{\alpha} R(t-x)}\frac{1}{\sqrt{2\pi R}}\sum_{l\in Z}a_{l+\alpha} e^{-i\tfrac lR(t-x)},\nn
\eea
where for concreteness we chose the fermions to be right-moving. The mode operators obey standard anti-commutation relations $\{a_r,a^\dagger_s\}=\delta_{rs}$, and 
the ground state $|\Omega\rangle$ is annihilated by $a_{r>0}$ and $a^{\dagger}_{r<0}$ operators.
For $\alpha = 1/2$ the fermions obey anti-periodic (or Neveu-Schwarz) boundary conditions while for $\alpha=0$ the fermions obey periodic (or Ramond) boundary conditions. The latter case needs to be treated separately due to the existence of a zero-mode \cite{Klich:2015ina}. 

For simplicity we will discuss only the case when $\alpha\neq 0$ and derive the entanglement Hamiltonian and entanglement entropy when the entangling region is an interval $V=(a, b)$, and the fermion system is in an excited state of definite positive momentum: $a^{\dagger}_{k+\alpha}|\Omega\rangle$.  The ground state $|\Omega\rangle$ is defined such that it is annihilated by all $a_{l+\alpha}$ with positive $l$: $a_{l+\alpha}|\Omega\rangle=0$ for $l\ge 0$. [Alternatively, instead of $a^{\dagger}_{k+\alpha}|\Omega\rangle$, with $k\geq 0$, a similar treatment is possible for the case when the excited state is a state of negative momentum: $a_{k+\alpha}|\Omega\rangle,$ with $k\leq 0$.]

For free systems without a degenerate ground state  all expectation values of operators are determined from the knowledge of the 2-point function using Wick's theorem. Then, the entanglement Hamiltonian $H_V$ on some entangling region $V$ is determined from the Green's function $G$ projected on the region $V$, $P_V G P_V$, as follows \cite{Peschel}: 
\bea
H_V=-\log((P_V G P_V)^{-1}-1).
\eea
The logarithm is evaluated with the help of the associated resolvent
\bea 
L(\beta)=(P_V G P_V-\tfrac 12+\beta)^{-1},\label{resgen}
\eea
via the integral
\bea
H_V=-\int_{\tfrac 12}^\infty d\beta \bigg(L(\beta)+L(-\beta)\bigg).
\eea

\begin{figure} 
\includegraphics[scale=1.]{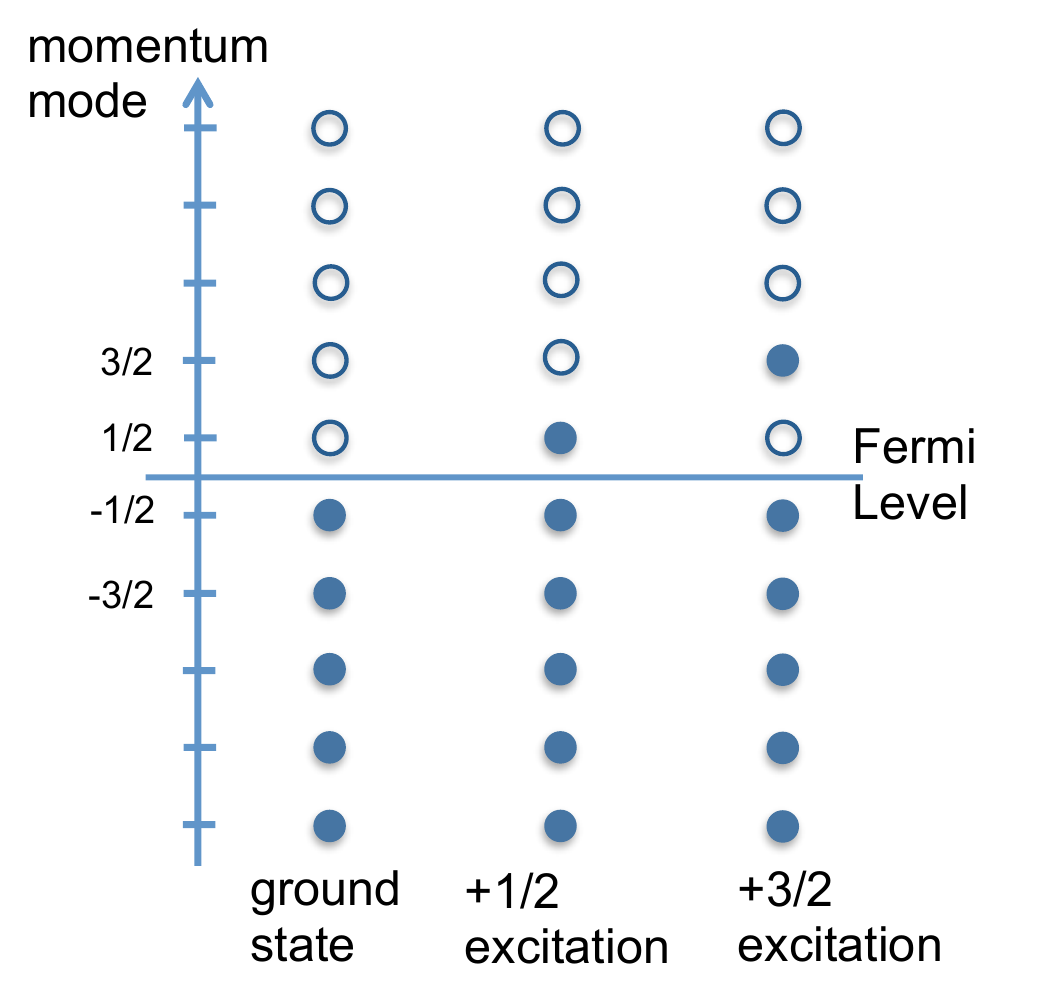}
\caption{States of the Dirac sea for anti-periodic wave functions. The $r=1/2$ excited state is equivalent to a unitary shift of the Green's function. We concentrate on the non-trivial $r=3/2$ excitation.}
\label{Excitedtates}
\end{figure}
We now focus on chiral Dirac fermions on a circle. Fig. As our excited state we choose adding a particle at momentum $r=3/2$. 

The Green's function evaluated on the excited state 
\begin{align}
G^{\alpha }_{(k)}(x,y)&= \braket{\Omega| a_{\alpha +k}  \Psi  (x+i0^+)\Psi^{\dagger}(y) a^{\dagger}_{k+\alpha} |\Omega}\nn
&=\frac{1}{2 \pi R} \sum_{r,q\in Z +\alpha } \braket{ \Omega |  a_{\alpha +k} a_{r} a^{\dagger }_{q} a^{\dagger }_{k+ \alpha}     |\Omega}   e^{\frac{i [r (x+i0^+)-q y]}{R}}\nn
&= \left(\frac{1}{2 \pi  R} \sum_{l=0}^\infty e^{ \frac{i (l+ \alpha ) [(x-y)+i0^+]}{R}} \right)
-\frac{1}{2 \pi  R}e^{\frac{i (k+\alpha ) (x-y)}{R}}\nn
&=e^{i\frac{\alpha} {R}(x-y)}n(x,y)-\frac{1}{2 \pi  R}e^{\frac{i (k+\alpha ) (x-y)}{R}}
\end{align}
is written in terms of the single-particle   projector onto positive (integer) momentum modes  
\bea
n(x,y)\equiv \frac{1}{2\pi R}\sum_{l=0}^{\infty}e^{i\tfrac lR (x-y+i0^+)}=\langle x|n|y\rangle,
\eea
plus a rank one perturbation:
\bea
G^{\alpha}_{(k)}=U_{\alpha} n U_{\alpha}^{-1}+\delta G^{\alpha}_{(k)}. \label{greensdirac}
\eea
In (\ref{greensdirac})  $U_{\alpha}$ are unitary operators which shift the integer momenta by 
$\alpha$: $U_{\alpha}|l\rangle=|l+\alpha\rangle$, where the single-particle momentum modes in position space are $\langle x|l\rangle=\frac{1}{\sqrt{2\pi R}}e^{i\tfrac {l}{R}x}$. The rank one perturbation in (\ref{greensdirac}) is given by
\begin{align}
\delta G^{\alpha }_{(k)}=-U_{\alpha} |k\rangle\langle  k|U_{\alpha}^{-1} .
\end{align} 
This is similar to the situation discussed in \cite{Klich:2015ina} where the Green's function for chiral Majorana periodic fermions in the ground state was written as a projector plus a rank one perturbation: $G=n-\tfrac 12 |0\rangle\langle0|$, with $|0\rangle$ the zero-momentum single particle state. The resolvent $(P_Vn P_V -\tfrac 12 + \beta -\tfrac 12 P_V|0\rangle\langle 0|P_V)^{-1}$ was derived by formally expanding the inverse, treating the zero-mode contribution as a perturbation, and subsequently resumming the series expansion. 
Here we apply the same trick. The first observation is that the resolvent for generic boundary conditions 
\bea
L^\alpha_{(k)}(\beta)=(P_V G^{\alpha}_{(k)} P_V -\tfrac 12+\beta)^{-1}
\eea
can be written as another unitary-transformed resolvent 
\bea
L^\alpha_{(k)}(\beta) &=& U_{\alpha}L_{(k)}(\beta)U_{\alpha}^{-1},\nn
L_{(k)}(\beta)&\equiv&(P_V n P_V - P_V|k\rangle\langle k|P_V-\tfrac 12+\beta)^{-1}.
\eea
Second, to compute  $L_{(k)}(\beta)$ we treat the term $-P_V|k\rangle\langle k|P_V$ as a Dyson perturbation, expand and then re-sum. The end result is   
\bea
(P_V n P_V - P_V |k\rangle\langle k|P_V-\tfrac 12+\beta)^{-1}=N(\beta) +\frac{N(\beta) P_V|k\rangle\langle k|P_V N(\beta)}{1-\langle k |P_V N(\beta) P_V|k\rangle},
\eea
where $N(\beta)$ is the resolvent associated with the projector $n$:
\bea
N(\beta)=(P_V n P_V -\tfrac 12+\beta)^{-1}.
\eea
$N(\beta)$ in turn is given by the solution to a Riemann-Hilbert problem as discussed in  \cite{Klich:2015ina} and 
revisited at length in Appendix \ref{apres}:
\bea
\langle x|P_V N(\beta)P_V |y\rangle=\Theta_V(x)\Theta_V(y)\bigg[\frac{\delta(x-y)}{\beta-\tfrac 12}-\frac{e^{-i h(Z^+(x)-Z^+(y))}}{\beta^2-\tfrac 14} n(x,y)\bigg],\label{Nn}
\eea
with 
\bea
h\equiv \frac{1}{2\pi}\ln\frac{\beta+\tfrac 12}{\beta-\tfrac 12},
\eea
and, for the case when then entangling region is a single interval $V=(a,b)$, the function $Z^+(x)$ is defined as\footnote{For a multiple interval formula, see Appendix \ref{apres}.}
\bea
Z^+(x)\equiv \ln\frac{e^{\frac iR(x+i0^+)}-e^{\frac iRa}}{e^{\frac iR(x+i0^+)}-e^{\frac iR b}}\equiv \ln z_+(x).
\eea
 $\Theta_V(x)$ is the characteristic function of the entangling region $V$: it is equal to 1 if $x\in V$ and 0 otherwise.

The conclusion of this analysis is that the resolvent for the excited state 
is given by the resolvent for the ground state, plus an extra term:
\begin{align}
\langle x|\delta L^{\alpha}_{(k)}(\beta)|y\rangle = e^{i\frac{\alpha}R(x-y)}
\frac{\braket{x| N(\beta) P_V |k} \braket{k|P_V N(\beta) |y}}{1-\braket{k|P_V N(\beta)P_V|k}}.
\end{align}

For the sake of concreteness we will consider the first excited state and take $k=1$
and compute the change to the resolvent for a {\it single interval} $V=(a,b)$. The resolvent matrix elements can be 
explicitly evaluated by contour integrals. We give a simpler algebraic derivation, in Appendix \ref{Apa}. Concretely, we find\footnote{For the case of multi-interval entangling regions, use instead the matrix elements computed in \ref{Apb}. }
%\begin{align}
%\braket{x|N(\beta)|k\text{=}1}&=\frac{1}{\beta- \tfrac{1}{2}} \frac{1}{\sqrt{2\pi R}}
%e^{- ihZ^+(x)} \bigg(ih(e^{\frac{i}{R}b }-e^{\frac{i}{R}a} ) +e^{\frac{i}{R}x}\bigg) \nn
%\braket{k\text{=}1|N(\beta)|k\text{=}1} &= 1-e^{\frac{h}{R}(a-b)} -4 h^{2}\sin^2(\frac{a-b}{2R}) 
%e^{\frac{h}R(a-b)} \end{align}
\bea
\braket{x|N(\beta)P_V|k\text{=}1}=\frac{\Theta_V(x)}{\beta-\tfrac 12}z_+^{-ih}(x) \frac{1}{\sqrt{2\pi R}}\bigg(e^{\tfrac iR x}+ih (e^{\tfrac iR b}-e^{\tfrac iR a})\bigg).\label{n1}
\eea
The other matrix element of interest, $\braket{k\text{=}1|P_V N(\beta)|x}$, is obtained from (\ref{n1}) by complex conjugation, given the hermiticity of the resolvent
\bea
\braket{k\text{=}1|P_V N(\beta)|x}=\frac{1}{\beta-\tfrac 12}\Theta_V(x) z_+^{ih}(x)e^{-2\pi h}e^{\tfrac hR(a-b)} \frac{1}{\sqrt{2\pi R}}\bigg(e^{-\tfrac iR x}-ih (e^{-\tfrac iR b}-e^{-\tfrac iR a})\bigg).
\nn
\label{n1c}
\eea
We also used that
\bea
\bigg[z_+^{-ih}(x)\bigg]^*=\bigg[\bigg(\frac{e^{\tfrac iR(x+i0^+)}-e^{\tfrac iR a}}{e^{\tfrac iR(x+i0^+)}-e^{\tfrac iR b}}\bigg)^{-ih} \bigg]^*=z_+^{ih}(x) e^{-2\pi h}e^{\tfrac hR(a-b)}.
\label{zpcc}\eea

Assembling the pieces, the  change in the resolvent relative to the ground state is:
\begin{align}
\langle x|\delta L^\alpha_{(1)}(\beta) |y\rangle = \Theta_V(x) \Theta_V(y)\frac{4 \sinh^{2}(\pi h)}{2\pi R} e^{\tfrac iR \alpha  (x-y)} z_+^{-ih}(x) z_+^{ih}(y)
 \frac{A h^2+ i h B(x,y)+ e^{\tfrac iR (x-y)} }{1+ A h^{2} },\label{resD1}
\end{align}
where
\begin{align}
A&=4 \sin^{2}( \frac{b-a}{2R}),\nn
B(x,y)&=e^{-\tfrac iR y}(e^{\tfrac iR b}-e^{\frac iR a})- e^{\tfrac iR x}(e^{-\tfrac iR b}-e^{-\tfrac iR a}).\label{ab}
\end{align}
This enables us to find the change in the entanglement Hamiltonian for chiral Dirac fermions in the first excited state, relative to the ground state: 
\begin{align}
\delta H_V^\alpha &=- \int_{1/2}^{\infty} d \beta  \bigg[\delta L^\alpha_{(1)}(\beta) + \delta L^\alpha_{(1)}(-\beta)\bigg].
\end{align} 
 In evaluating the integral it is useful to change the integration variable from $\beta$ to $h$, which leads to 
\begin{align}
\braket{x|\delta H_V^\alpha|y}&=
-\int_{-\infty}^\infty dh\frac{2\pi}{4\sinh^2(\pi h)} \braket{x|\delta L^\alpha_{(1)}(h)|y} 
\nn &=-2\pi\frac{e^{\frac{i}R \alpha (x-y)}}{R}\Theta_V(x)\Theta_V(y)\bigg[ \delta (Z(y)-Z(x))\nn&+ 
 \frac{1}{2A}\Theta(Z(y)-Z(x)) e^{\frac {Z(x)-Z(y)}{\sqrt{A}}}
\bigg( -B(x,y)+\sqrt{A}(e^{\frac iR (x-y)}-1)\bigg)
\nn
&+ \frac{1}{2A}\Theta(Z(x)-Z(y))e^{\frac {Z(y)-Z(x)}{\sqrt{A}}}\bigg(B(X,y)+\sqrt{A}(e^{\tfrac iR (x-y)}-1) \bigg)
\bigg],\label{EHDirac}
\end{align}
and where $A, B(x,y)$ were previously defined in (\ref{ab}). The entanglement Hamiltonian in (\ref{EHDirac}) has a delta-function term whose interpretation is that of a chemical potential times the inverse entanglement temperature $\beta(x)=2\pi |Z'(x)|$. The remaining terms in (\ref{EHDirac}) are exponentially falling off with the distance $|Z(y)-Z(x)|$. For the single interval entangling region, $Z(x)$ is monotonic and  $\Theta (Z(x)-Z(y))$ equals $\Theta(x-y)$.

The entanglement entropy 
\bea
S_V=-\rm{Tr}_V \rho_V\ln\rho_V
\eea
can also be expressed in terms of the resolvent  (\ref{resgen})
\bea 
S&=&-\text{Tr}_V\,\bigg[G \ln G - (G-1)\ln(G-1)\bigg]\nn
&=&-\int_{1/2}^\infty d\beta\,\text{Tr}_V\bigg[(\beta-\tfrac 12)\bigg(L(\beta)-L(-\beta)\bigg)-\frac{2\beta}{\beta+\tfrac 12}\bigg].
\eea

\begin{figure} 
\includegraphics[scale=0.6]{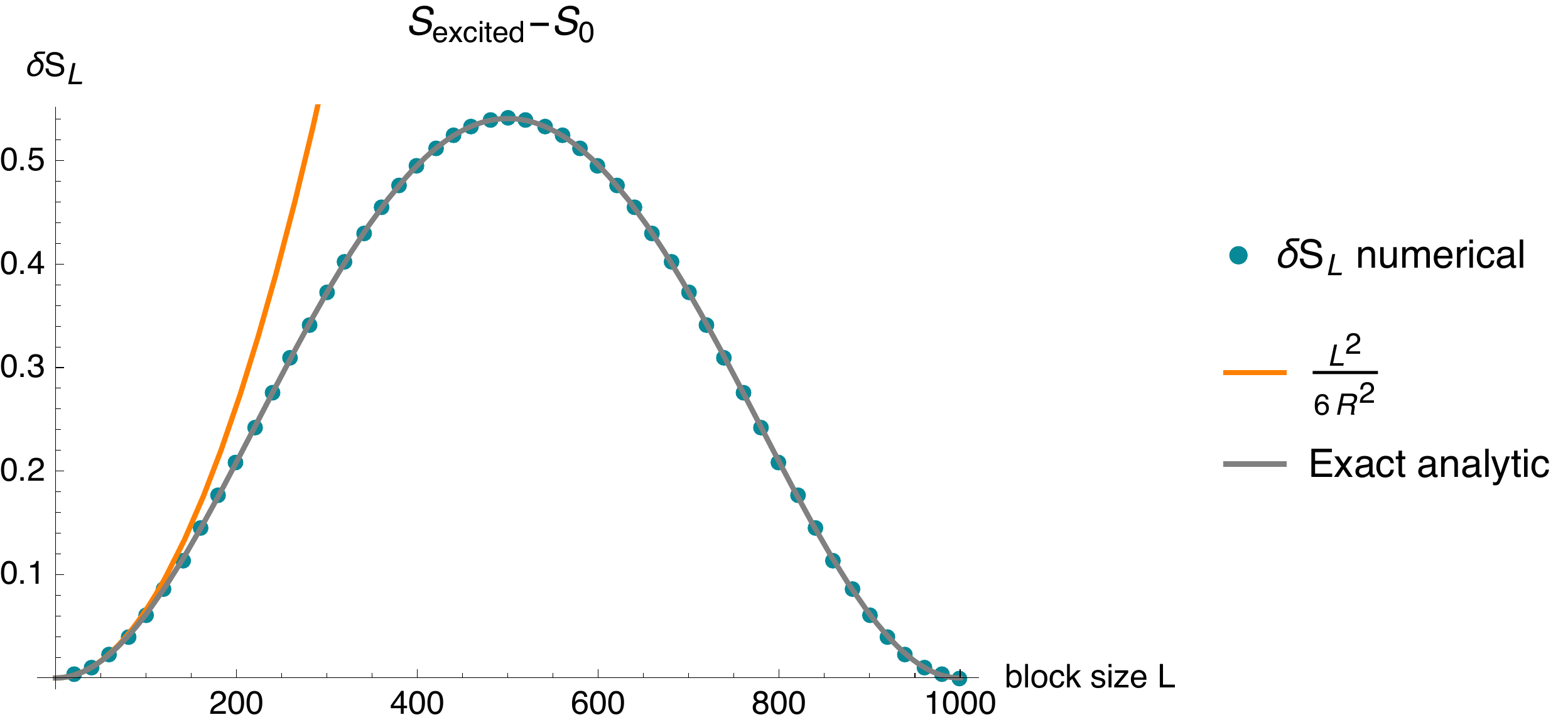}
\caption{Beyond first law behavior: $\delta S$ for an extra particle at the $k=1$ (momentum $\frac{3}{2R}$) excited state, compared with our formula \eqref{delta sv}, for fermions on a ring with 1000 sites.}
\label{ExcitedEntropyNumericalExact}
\end{figure}

Using (\ref{resD1}), the change in the entanglement entropy of the chiral Dirac fermions in the first excited state relative to the ground state
\bea
\delta S_V =-\int_{1/2}^\infty d\beta\,(\beta-\tfrac 12)\text{Tr}_V 
\bigg[ \delta L^\alpha_{(1)} (\beta) - \delta L^\alpha_{(1)} (-\beta) \bigg]
\eea 
evaluates to 
\begin{align}
\delta S_V
&=\int_{0}^{\infty} d h \, \text{Tr} \bigg[ \delta L^\alpha_{(1)} (h) - \delta L^\alpha_{(1)} (-h) \bigg] \frac{2 \pi (1- \coth(\pi h))}{8 \sinh^{2} (\pi h)}\nn
&= 
-\int_{0}^{\infty} d h\, \frac{   2 A h (1-\coth (\pi  h))}{1+A h^2}\nn
&=-2\bigg[\sin(\frac{L}{2R})+\log(2\sin(\frac{L}{2R}))+\psi(\frac{1}{2\sin(\frac{L}{2R})})\bigg],\label{delta sv}
\end{align}  
where $\psi(z)=\frac{d}{dz}\ln\Gamma(z)$ and $L=b-a$. Figure \ref{ExcitedEntropyNumericalExact} shows the remarkable fit between our exact formula \eqref{delta sv} computed in the continuum with numerical results for fermions on a discretized ring where negative momentum states have been filled.

Of particular interest is the limit $L/R \to 0$. Here, for any finite $L$, as $R\to\infty$, we expect the change in entropy due to an additional delocalized particle to vanish together with its effect on energy density and particle density.
%As a way of checking the expression for the correction to the ground state entanglement entropy obtained above in (\ref{delta sv}), we can take the limit when $R\to \infty$
{}\footnote{We don't need the exact final expression in (\ref{delta sv}) to compute the large $R$ limit of the entanglement entropy correction. 
In this limit, the integrand can be approximated by $-2Ah(1-\coth(\pi h))$, since the denominator contribution is cut off by the exponential fall off of the numerator. The integral is then readily evaluated to $1/6\, (L/R)^2$.}:
\bea 
\delta S_V \simeq \frac{L^2}{6R^2}.\label{eelargeRDirac}
\eea 
\subsection{Rank two perturbations: chiral Majorana fermions}\label{subsec:MajoranaExcited}
%%%%%%%%%%%%%%%%%%%%%%%%%%%%%%%%%%%%%%%%%%%%%%%%%%%%%%%%%%%%%%%%%%%%%%%%%%

 The Green's function for anti-periodic chiral Majorana fermions\footnote{We exclude for now the case of periodic Majorana fermions since that has the added difficulty of accounting for the zero mode contribution. Here the ground state $|\Omega\rangle$ is defined such that $b_r|\Omega\rangle=0$ for $r>0$, and the mode expansion operators obey the anti-commutation relations $\{b_r, b_s\}=\delta_{r+s}$. } on the excited state $b_{-(k+\tfrac 12)}|\Omega\rangle$ with $k\geq 0$ is given by
\begin{align}
G(x, y)=&\braket{\Omega |   b_{k+\tfrac 12}  \psi (x+i \epsilon) \psi ( y)  b_{-(k+\tfrac 12)}   |\Omega } \nn=&
\frac{1}{2 \pi  R}\sum _{r,q\in  Z+\tfrac 12}  \braket{ \Omega |b_{k+\tfrac 12}  b_q b_r  b_{-(k+\tfrac 12)}|\Omega }   e^{\frac iR [q y+r (x+i0^+)]}\nn
=&\bigg(\sum_{l=0}^{\infty}  \frac{1}{2 \pi  R} e^{ \frac iR (\frac 12 +l) [(x-y)+i0^+]} \bigg)+\frac{e^{\frac iR (\frac 12 +k) (x-y)}}{2 \pi  R}-\frac{e^{-\frac iR (\tfrac 12 +k) (x-y)}}{2 \pi  R},
\end{align}
and can be expressed as a rank 2 perturbation of an integrable operator:
\bea 
G=U_{\alpha=\tfrac 12}\bigg( n +|k\rangle\langle k|- |\text{-}(k+1)\rangle\langle\text{-}(k+1)|\bigg)U_{\alpha=\tfrac 12}^{-1}.
\eea

To allow for a bit more generality we discuss next how to construct the resolvent 
\bea
L(\beta)=(P_V n P_V + c_1P_V|k_1\rangle \langle k_1|P_V+ c_2 P_V|k_2\rangle \langle k_2|P_V +\beta-\tfrac 12)^{-1},\label{r2p}
\eea
where $c_1$ and $c_2$ are real-valued numbers.
As in the case of the rank one perturbation, we factor out the resolvent $N(\beta)=(P_V n P_V+\beta-\tfrac 12)^{-1}$. We expand $(1+N(\beta)  P_V ( c_1|k_1\rangle \langle k_1|+ c_2|k_2\rangle \langle k_2|)P_V)^{-1}$ and then we re-sum. The re-summation can be performed straightforwardly  after the following treatment. First we define new states
\bea
&&|l_1\rangle= |k_1\rangle\ + L_1 |k_2\rangle\nn
&&|l_2\rangle=|k_1\rangle+ L_2 |k_2\rangle,
\eea
where the coefficients $L_1$ and $L_2$ are determined such that 

i) the sum over the $|k_1\rangle$ and $|k_2\rangle$-projectors (namely 
$c_1 |k_1\rangle\langle k_1| +c_2 |k_2\rangle\langle k_2|)$ can be written as a sum over $|l_1\rangle$ and $|l_2\rangle$-projectors; this holds if
\bea
c_1 L_1 L_2^*+c_2=0\label{cond1}
\eea

ii) the matrix elements $\langle l_1|P_V N(\beta)P_V |l_2\rangle$ vanish:
\bea
\langle k_1|P_V N(\beta) P_V|k_1\rangle+L_1^* \langle k_2|P_V N(\beta) P_V|k_1\rangle+L_2 \langle k_1|P_V N(\beta)|P_V k_2\rangle+ L_1^* L_2 \langle k_2|P_V N(\beta)P_V |k_2\rangle=0.\label{cond2}\nn
\eea

From (\ref{cond1}) and (\ref{cond2}) we identify the coefficients $L_1, L_2$.
Then the computation  of the inverse of the operator
\bea
&&1+N(\beta)P_V (c_1|k_1\rangle \langle k_1|+ c_2|k_2\rangle \langle k_2|)P_V = 1+ N(\beta)\,P_V\bigg(\frac{c_1 |L_2|^2 +c_2}{|L_1-L_2|^2} |l_1\rangle \langle l_1|+\frac{c_1 |L_1|^2 +c_2}{|L_1-L_2|^2} |l_2\rangle \langle l_2|\bigg)P_V\nn
%&&=1+N(\beta)\frac{|L_2|^2 c_1 |l_1\rangle\langle l_1|+c_2 |l_2\rangle\langle l_2|}{\frac{c_2}{c_1}+|L_2|^2},
\eea
proceeds just as for the case of the rank one perturbation and yields
\bea
\bigg(1+N(\beta)P_V(c_1|k_1\rangle \langle k_1|+ c_2|k_2\rangle \langle k_2|)P_V \bigg)^{-1}&=&1 - \frac{c_1 |L_2|^2 +c_2}{|L_1-L_2|^2}\frac{N(\beta)P_V |l_1\rangle\langle l_1|P_V}{1-\frac{c_1 |L_2|^2 +c_2}{|L_1-L_2|^2}\langle l_1|P_V N(\beta)P_V|l_1\rangle}\nn
&&-\frac{c_1 |L_1|^2 +c_2}{|L_1-L_2|^2}\frac{N(\beta)P_V |l_2\rangle \langle l_2|P_V }{1-\frac{c_1 |L_1|^2 +c_2}{|L_1-L_2|^2}\langle l_2|P_V N(\beta)P_V |l_2\rangle}.\nn
\eea

Putting everything together, the resolvent in (\ref{r2p}) is given by
\bea
\langle x| L(\beta)|y\rangle &=&\langle x| N(\beta)|y\rangle -\frac{c_1 |L_2|^2 +c_2}{|L_1-L_2|^2}\frac{\langle x |N(\beta)P_V |l_1\rangle\langle l_1|P_V N(\beta)|y\rangle}{1-\frac{c_1 |L_2|^2 +c_2}{|L_1-L_2|^2}\langle l_1|P_V N(\beta)P_V |l_1\rangle}\nn
&&-\frac{c_1 |L_1|^2 +c_2}{|L_1-L_2|^2}\frac{\langle x| N(\beta) P_V |l_2\rangle \langle l_2|P_V N(\beta)|y\rangle}{1-\frac{c_1 |L_1|^2 +c_2}{|L_1-L_2|^2}\langle l_2|P_V N(\beta)P_V |l_2\rangle}.\nn
\eea

Let's consider next a concrete case, the first excited state of a chiral, anti-periodic Majorana fermion with $k=0$. 
Then, $|k_1\rangle=|k=0\rangle$, $|k_2\rangle=|k\text{=-}1\rangle$, $c_1=1$ and $c_2=-1$. From (\ref{cond1}) we find $L_2^*=(L_1)^{-1}$. Substituting in (\ref{cond2}) leads to
\bea
&&\langle k=0|P_V N(\beta)P_V |k=0\rangle+\langle k\text{=-}1|P_V N(\beta)P_V | k\text{=-}1\rangle
\nn&&+\frac{1}{L_2}\langle  k\text{=-}1|P_V N(\beta)P_V |k=0\rangle + L_2
\langle k=0|P_V N(\beta)P_V |k\text{=-}1\rangle=0,
\eea
and finally to the sought-after resolvent
\bea
\langle x| L_{\text {excited}}(\beta)|y\rangle &=&\langle x| N(\beta)|y\rangle \nn
&&+
\frac{1+N_{-1,-1}}{(1-N_{0,0})(1+N_{-1,-1})+N_{0,-1}N_{-1,0}}
\langle x|N(\beta)P_V |k=0\rangle\langle k=0|P_V N(\beta)|y\rangle\nn
&&-
\frac{1-N_{0,0}}{(1-N_{0,0})(1+N_{-1,-1})+N_{1,-1}N_{-1,0}}
\langle x|N(\beta)P_V |k=\text{-}1\rangle\langle k=\text{-}1|P_V N(\beta)|y\rangle\nn
&&-\frac{N_{0,-1}}{(1-N_{0,0})(1+N_{-1,-1})+N_{0,-1}N_{-1,0}}
\langle x|N(\beta)P_V |k=0\rangle\langle k=\text{-}1|P_V N(\beta)|y\rangle\nn
&&-\frac{N_{-1,0}}{(1-N_{0,0})(1+N_{-1,-1})+N_{0,-1}N_{-1,0}}
\langle x|N(\beta)P_V |k=\text{-}1\rangle\langle k=0|P_V N(\beta)|y\rangle,\nn\label{resolventMaj}
\eea
where the matrix elements $N_{0,0}=\braket{k=0|P_V N(\beta)P_V |k=0},$ $N_{0,-1}=
\braket{k=0|P_V N(\beta)P_V |k=\text{-}1},$ etc. are computed in Appendix \ref{Apa}. As in the previous section, from
\bea
&\langle x|N(\beta) P_V|k\text{=-}1\rangle = \frac{1}{\beta-\tfrac 12}\Theta_V(x) X_+^{-1}(x) \frac{e^{-i\tfrac xR}}{\sqrt{2 \pi R}} e^{\tfrac hR(b-a) }, \label{n-1}\eea 
using the hermiticity of the resolvent and (\ref{zpcc}) we can determine its complex conjugate
\bea
&\langle k\text{=-}1|N(\beta) P_V|x\rangle=\frac{1}{\beta-\tfrac 12}\Theta_V(x) X_+(x) \frac{e^{i\tfrac xR}}{\sqrt{2 \pi R}} e^{-2\pi h },
\label{n-1c}
\eea  
For a single interval $V=(a,a+L)$, substituting the appropriate matrix elements into (\ref{resolventMaj})
the resolvent of chiral Majorana fermions in the first excited state takes the form
\bea 
\braket{x|L_{\text {excited}}(\beta)|y}&=&\braket{x|N(\beta)|y}\nn&& + 
z_+^{-ih}(x)z_+^{ih}(y)\Theta_V(x)\Theta_V(y)\bigg(\frac{
1 - e^{\tfrac iR(x-y)} -ihB(y,x) }{1+4 h^2 \sin ^2(\frac{L}{2 R})}\bigg)
 \frac{4\sinh^2(\pi h)}{2\pi R}.\label{resolventMaj1}\nn
\eea 
The first excited state entanglement Hamiltonian,  relative to the ground state, is
\bea 
\braket{x|\delta H_V|y} &=&-\frac 12 \int_{-\infty}^\infty dh\,\frac{2\pi}{4\sinh^2(\pi h)}
\braket{x|\delta L^{\alpha=1/2}_{\text {excited}}(h)|y},\label{EHMaja}
\eea 
where the fact that we're dealing with real (Majorana) fermions is responsible for the extra factor of $(-1/2)$ on the right hand side of (\ref{EHMaja}) relative to the previously discussed case which involved Dirac fermions\footnote{For more details see \cite{Klich:2015ina}.}. 
The $h$-integral in (\ref{EHMaja}) is essentially the Fourier-transform of a rational function equal to the ratio of a polynomial of degree 1 and a polynomial of degree 2 in $h$. The Fourier-transform which gives the entanglement Hamiltonian can be performed using Cauchy's theorem after partial fractioning, leading to 
\bea 
\braket{x|\delta H_V|y} &=&
-\frac{\pi}{2A} \frac{e^{\frac{i}{2R} (x-y)}}{R}
\bigg[\Theta(Z(x)-Z(y))e^{\frac {Z(y)-Z(x)}{\sqrt{A}}} \bigg( -B(y,x)+\sqrt{A}(1-e^{-\frac iR (x-y)})\bigg)
\nn
&&+\Theta(Z(x)-Z(y))e^{\frac {Z(y)-Z(x)}{\sqrt{A}}}\bigg(B(x,y)+\sqrt{A}(1-e^{-\tfrac iR (x-y)}) \bigg)
\bigg]\Theta_V(x)\Theta_V(y)
,\nn
\eea
where  we recall that $A, B(x,y)$ were previously defined in (\ref{ab}). 
 All the terms are falling off exponentially with the distance $|Z(y)-Z(x)|$. As opposed to the chiral Dirac fermion case analyzed previously the term proportional to a delta-function is absent, which is to be expected, since we are dealing with real fermions.
 
 The entanglement entropy correction of the Majorana NS fermions in their first excited state relative to the ground state has the form
 \bea
\delta S_V =\tfrac 12 \int_{1/2}^\infty d\beta\,(\beta-\tfrac 12)\text{Tr}_V 
\bigg[ \delta L^{\alpha=1/2}_{\text {excited}} (\beta) - \delta L^{\alpha=1/2}_{\text {excited}} (-\beta) \bigg]
\eea 
evaluates to 
\begin{align}
\delta S_V
&=-\tfrac 12\int_{0}^{\infty} d h \, \text{Tr}_V \bigg[ \delta L^{\alpha=1/2}_{\text{excited}} (h) - \delta L^{\alpha=1/2}_{\text{excited}} (-h) \bigg] \frac{2 \pi (1- \coth(\pi h))}{8 \sinh^{2} (\pi h)}\nn
&= 
-\int_{0}^{\infty} d h\, \frac{    A h (1-\coth (\pi  h))}{1+A h^2}\nn
&=-\bigg[\sin(\frac{L}{2R})+\log(2\sin(\frac{L}{2R}))+\psi(\frac{1}{2\sin(\frac{L}{2R})})\bigg],\label{delta sv}
\end{align}  
 and, in the large radius limit it is approximated by
 \bea 
 \delta S_V&\simeq&\frac 12 \frac{L^2}{6R^2}.
 \eea

%%%%%%%%%%%%%%%%%%%%%%%%%%%%%%%%%%%%%%%%%%%%%%%%%%%%%%%%%%%%%%%%%%%%%%%%%%%%
\section{The entanglement entropy of low temperature, chiral NS Dirac fermions}\label{sec:DiracLowTemp}
%%%%%%%%%%%%%%%%%%%%%%%%%%%%%%%%%%%%%%%%%%%%%%%%%%%%%%%%%%%%%%%%%%%%%%%%%%%%

In this section we give one more example based on a rank 2 perturbation of an integrable operator. 
Consider the finite temperature expansion of the Green's function of chiral NS Dirac fermions
\begin{align}
G(x,y)&=\frac{\text{Tr}\bigg(e^{-\tfrac1{T} H} \Psi(x)\Psi^\dagger(y)\bigg)}{\text{Tr}(e^{-\tfrac1{T} H})}\nn
&=\frac1{\text{Tr}(e^{-\tfrac1{T} H})}\bigg[\braket{\Omega|\Psi(x)\Psi^{\dagger}(y)|\Omega} +e^{-\tfrac1{2TR} } \langle\Omega| a_{r=1/2}^{} \Psi(x)\Psi^{\dagger}(y)a^\dagger_{r=1/2}|\Omega\rangle\nn&+ e^{-\tfrac{1}{2TR}} \langle\Omega| a^\dagger_{r=-1/2} \Psi(x)\Psi^{\dagger}(y)a^{}_{r=-1/2}|\Omega\rangle
+\dots \bigg]
\end{align}
where we've used the Hamiltonian
\bea 
H=\sum_{r>0} r a^\dagger_r a_r +\sum_{r<0} r a_r a^\dagger_r. 
\eea
Then, to first order in a low temperature expansion, the Green's function takes the form
\bea
G\simeq U_{\alpha=1/2}\bigg( n - e^{-\tfrac1{TR}} (\ket{k\text{=}0}\bra{k\text{=}0}-\ket{k\text{=-}1}\bra{k\text{=-}1} )\bigg)U_{\alpha=1/2}^{-1},\nn
\eea
where $T$ is the temperature and we truncated the series to lowest order in $e^{-\tfrac 1{2TR}}$.

When restricted to a subset $V$, and to lowest order in the temperature, this is a rank two perturbation of the integrable operator  $ P_{V} n P_{V} $.  However, we have to work consistently to first order in the expansion parameter $e^{-\tfrac 1{2TR}}$.
Writing 
\bea
\delta G \equiv - e^{-\frac1{2TR}} \bigg(\ket{k\text{=}0}\bra{k\text{=}0}-\ket{k\text{=-}1}\bra{k\text{=-}1}\bigg)
\eea
we note that to linear order in   $\delta G$, the finite temperature resolvent is 
\bea
N_T(\beta) &\equiv& \bigg( P_V\,n\,P_V + P_V\,\delta G \,P_V+ \beta -1/2\bigg)^{-1}  \nn
&\simeq& N(\beta)- 
N(\beta) \,P_V\,\delta G \, P_V\,N(\beta).
\eea
We are interested in  the first order finite temperature corrections to the entanglement entropy
\bea
S_V
&=&-\int_{1/2}^\infty d\beta\,\text{Tr}_V\bigg[(\beta-\tfrac 12)\bigg(N_T(\beta)-N_T(-\beta)\bigg)-\frac{2\beta}{\beta+\tfrac 12}\bigg]\nn
&\simeq&-\int_{1/2}^\infty d\beta\,\text{Tr}_V\bigg[
(\beta-\tfrac 12)\bigg(N(\beta)-N(-\beta)\bigg)-\frac{2\beta}{\beta+\tfrac 12}
\bigg]\nn
&&+\int_{1/2}^{\infty}d\beta\,\text{Tr}_V\bigg[(\beta-\tfrac 12)\bigg(N(\beta)P_V\,\delta G\,P_V N(\beta)-N(-\beta)P_V\,\delta G\,P_V N(-\beta)\bigg)\bigg]. \label{eefintemp}
\eea

For a single interval $V=(a,b)=(a,a+L)$ entangling region we have:
\bea
&&\!\!\!\!\!\!\!\!\!\int_{1/2}^{\infty}d\beta(\beta-\tfrac 12)\text{Tr}_V
\bigg[N(\beta)P_V |k\text{=-}1\rangle\langle k\text{=-}1|P_V N(\beta)-
N(-\beta)P_V |k\text{=-}1\rangle\langle k\text{=-}1|N(-\beta)\bigg]\nn
&&\!\!\!\!\!\!\!\!\!=-2\,\frac L{R}\int_{0}^\infty dh (1-\coth(\pi h))\sinh(\frac hR L)=\bigg(1-
\frac {L}{2R}\cot\frac{L}{2R}\bigg),
\eea 
where we've substituted the resolvent matrix elements $\braket{x|N(\beta)|k=\text{-1}}$ computed in Appendix \ref{Apa}, so that 
\bea
\braket{x|N(\beta)|k=\text{-1}}\braket{k=\text{-1}|N(\beta)|x}=\Theta_V(x)
\frac{e^{-2\pi h}}{(\beta-\tfrac 12)^2} \frac{e^{\frac hR L}}{2\pi R},
\eea
and carried out $\text{Tr}_V$ by integrating over $x\in V$. Similarly, with results from Appendix \ref{Apa}
\bea 
\braket{x|N(\beta)|k=0}\braket{k=0|N(\beta)|x}=\Theta_V(x)
\frac{e^{-2\pi h}}{(\beta-\tfrac 12)^2} \frac{e^{-\frac hR L}}{2\pi R}
\eea 
we find
\bea
&&\!\!\!\!\!\!\int_{1/2}^{\infty}d\beta(\beta-\tfrac 12)\text{Tr}_V
\bigg[N(\beta)P_V |k\text{=}0\rangle\langle k\text{=}0|P_V N(\beta)-
N(-\beta)P_V |k\text{=}0\rangle\langle k\text{=}0|P_V N(-\beta)\bigg]\nn
&&\!\!\!\!\!\!=2\,\frac L{R}\int_{0}^\infty dh (1-\coth(\pi h))\sinh(\frac hR L)=-\bigg(1-
\frac {L}{2R}\cot\frac{L}{2R}\bigg).
\eea

Substituting these into (\ref{eefintemp}) we arrive at the first order temperature correction of the single interval entanglement entropy
\bea 
\delta S_V=2e^{-\frac1{2TR}} \bigg (1-\frac {L}{2R} \cot\frac{L}{2R}\bigg). 
\label{eefintemp1}
\eea 
This matches the first order finite temperature corrections obtained by Herzog and Cardy in \cite{Cardy:2014jwa}, if we take into account the fact that we are considering {\it chiral} Dirac fermions, so the entanglement entropy is half of the Dirac fermion result in \cite{Cardy:2014jwa}.

However, with our method we can just as easily study an entangling region which is a disjoint sum of several intervals, $V=\bigcup_i (a_i, b_i)$. Denoting by $L$ the total length of $V$, 
\bea
L=\sum_i(b_i-a_i),
\eea 
and using the matrix elements of the resolvent computed in Appendix \ref{Apb}, we conclude that
 the first order temperature correction to the entanglement entropy  
is still given by (\ref{eefintemp1}).
%%%%%%%%%%%%%%%%%%%%%%%%%%%%%%%%%%%%%%%%%%%%%%%%%%%%%%%%%%%%%%%%%%
\\
\section*{Acknowledgments} %We would like to thank ....  
The work of IK was supported in part by the NSF grant DMR-1508245.The work of DV and GW was supported in part by the DOE grant DE-SC0007894.

\section*{Appendix}
\begin{appendix}
\section{ The projected resolvent and the Riemann-Hilbert problem }\label{apres}%\label{sec:AppendixRHmethod}

Let $\mathcal{H}_{S^{1}}$ denote a quantum mechanical Hilbert space on a circle with angular coordinate $x \sim x + 2 \pi R $. It will be useful to treat $S^{1}$ as the circle 
\bea
\{z \in \mathbb{C};z=e^{\frac{ i x}{R}}\}.
\eea
 on the complex $z$ plane.  
Here we review the derivation of the resolvent
\bea\label{N}
(1+K)^{-1}.
\eea
where $K$ is an integrable operator of the form
\bea
K= \sum_{i=1}^{m} f_{i} n g_{i}, 
\eea 
where $n$ is a projector onto semi-positive modes $e^{i n x }$ with $\, n\geq 0$ and $f_{i}$ and $g_{i}$ are operators that are diagonal in position space.    If we extend the mode expansion on $S^{1}$ to a Laurent expansion on $\mathbb{C}$ then functions containing only positive/negative modes corresponds those functions that are analytic inside/outside the circle. 
For convenience we define the $m \times m$ matrices of operators $N$,$F$,$G$ with matrix elements 
\begin{align}
N_{ij} = n(x,y) \, \delta_{ij} ,\nn
F_{ij}= f_{j}(x) \delta_{i1} ,\nn
G_{ij}=g_{i}(x).
 \delta_{j1} 
 \end{align}
This effectively attaches an $m \times m$ dimensional internal space ${\cal I}$ to each point $x$, so these operators act on an enlarged Hilbert space 
\bea
\mathcal{H} = {\cal I} \otimes \mathcal{H}_{S^{1}}.
\eea 
We will show that 
\begin{align} \label{R}
(1+K)^{-1} &= [\mathbf{1} - FX_{+}^{-1}NX_{-}G]^{-1}_{11}\nn
  &= 1- \sum_{i,j,l} f_{i}  [X_+^{-1}]_{ij} n [X_{-}]_{jl}  g_{l} .
\end{align} 
where $X_{\pm}$ is the solution to the non-abelian Riemann Hilbert problem 
\bea\label{RHP} 
X=X_{-}^{-1}X_{+}.
\eea
with $X$ the $k \times k$ jump matrix defined on $S^{1}$
\bea \label{X}
X=\mathbf{1} +GF\nn
X_{ij}= \delta_{ij} +g_{i}f_{j}.
\eea
 The matrix elements of $X_{\pm}$ are diagonal in position space and must be analytic inside/outside of the curve $S^{1}$.  Equivalently, we can characterize $X_{\pm}$ as containing only (semi)positive or (semi) negative modes.  

To prove \eqref{R} consider the identity
\begin{align}\label{resolvent}
(1+K)^{-1} &= ([\mathbf{1} + FNG]_{11})^{-1} \nn
&= [(\mathbf{1} + FNG)^{-1}]_{11}.
\end{align}
Note that in the first equality the inverse is taken on $\mathcal{H}_{S^{1}}$ while in the second line the inverse is taken in $\mathcal{H}$. To proceed we used the identity 
\begin{align} 
(\mathbf{1} + FNG)^{-1}&= \mathbf{1}- FNG(\mathbf{1}+FNG)^{-1}\nn
&= \mathbf{1}- FN (\mathbf{1}+GFN)^{-1} G,
\end{align}
where in the second line we used the relation $ (1+AB)^{-1} A =A (1+BA)^{-1} $ with $ A=G$ and $B=FN$.
Thus the problem reduces to finding $( \mathbf{1}+GFN)^{-1}$.
In terms of the operator $X$ we can write
\begin{align}
\mathbf{1}+GFN = N_{\perp} + X N,
\end{align} 
where we defined 
\bea
N_\perp\equiv \mathbf{1}-N.
\eea
The Riemann-Hilbert problem now enters as follows.
Given a solution to the  Riemann-Hilbert problem \eqref{RHP}, the definition of $N$ as a projector onto semi-positive modes imply the crucial identities
\bea
N_{\perp}X_{-}^{-1} N_{\perp} = N X_{-}^{-1},\nn
NX_{+}^{-1} N =X_{+}^{-1} N.
\eea 
From these identities it can be checked that 
\bea \label{R^-1}
( \mathbf{1}+GFN)^{-1}= N X_{+}^{-1} N X_{-} + N_{\perp} X_{-}^{-1} N_{\perp} X_{-} .
\eea
Substituting this back into \eqref{resolvent}, gives the expression \eqref{R} for the resolvent.
For our applications to computations of the entanglement Hamiltonian,  we need the resolvent as a function of an auxiliary parameter $\beta$: 
\bea\label{N beta Res}
N(\beta )\equiv \bigg(K+ \beta  -\tfrac{1}{2}\bigg)^{-1}.\label{n0}
\eea
In this case the jump matrix is modified to
\bea
X= \mathbf{1} +\frac{1}{\beta-1/2} GF
\eea
and $N(\beta)$ takes the form 
\bea
N(\beta)= \frac{1}{\beta -\frac{1}{2}} 
\bigg(1- \frac{1}{\beta -\frac{1}{2}} \sum_{i,j,k} f_{i}  [X_+^{-1}]_{ij} n[X_{-}]_{jk}  g_{k} \bigg).
\eea
For the purpose of this paper, we are interested in the inverse
\bea
N(\beta)=(P_v n P_V + \beta-\frac 12)^{-1} = \frac{1}{\beta-\tfrac 12}\bigg(
\frac{1}{\beta-\tfrac 12} P_V \,n\, P_v +1\bigg)^{-1}.
\eea 
Then the previous analysis simplifies to a scalar Riemann-Hilbert problem. There is only one operator $f$ and one operator $g$, which are equal the projector on the entangling region $V$
\bea
f=g=P_V, 
\eea
and the we need the solution to the Riemann-Hilbert problem with the jump function
\bea
X=X_-^{-1}X_+=1 + \frac{1}{\beta-\tfrac 12}P_V.
\eea
For a region $V$ which is the sum of $N$ disjoint intervals on a circle of radius $R$, $V=\bigcup_{j=1}^N (a_j,b_j)$, the solution to the Riemann-Hilbert problem is readily available \cite{Klich:2015ina}
\bea
X_\pm(x)=\bigg(\prod_{j=1}^N\frac{e^{\tfrac iR(x\pm i0^+)}-e^{\tfrac iR a_j}}{e^{\tfrac iR(x\pm i0^+)}-e^{\tfrac iR b_j}}\bigg)^{ih}.\label{xpmninterv}
\eea

%%%%%%%%%%%%%%%%%%%%%%%%%%%%%%%%%%%%%%%%%%%%%%%%%%%%%%%%%%%%%%%%%%%%
\section{How to evaluate algebraically the resolvent  matrix elements}\label{sec:AppendixMatrixElements}

\subsection{Single interval entangling region}\label{Apa}

To see how we can evaluate matrix elements of the resolvent (\ref{Nn}) of the kind $\langle x|N(\beta)P_V|k\rangle$ and $\langle k|P_V N(\beta)|k'\rangle$ algebraically (that is, without the hassle of evaluating  complex integrals in the presence of cuts), let's start with the resolvent projected onto the region $V$  written as 
\bea
(\beta-\tfrac 12) N(\beta)\, P_V=(\beta-\tfrac 12)P_V N(\beta)=P_V-\frac{1}{\beta-\tfrac 12} P_V X_+^{-1} n X_- P_V, \label{Nn1}
\eea
where, as discussed in Appendix \ref{apres},  $\langle x|X_\pm|y\rangle $ are diagonal holomorphic matrices inside/outside the unit circle such that\footnote{
Since $P_V^2=P_V$, the equation (\ref{pv1}) can also be written as $\frac{\beta+\tfrac 12}{\beta-\tfrac 12}P_V=X_-^{-1} X_+ P_V$, or $X_-P_V=\frac{\beta-\tfrac 12}{\beta+\tfrac 12} X_+ P_V=e^{-2\pi h} X_+ P_V$. Also note that since all three operators are diagonal in position space they commute with one another.

}
\bea
\frac{1}{\beta-\tfrac 12} P_V =  X_-^{-1} X_+ -1. \label{pv1}
\eea
Moreover, in terms of previously introduced notation,
\bea
X_+(x)=z_+^{ih}(x), \qquad X_-(x)=z_-^{ih}(x).
\eea
Substituting (\ref{pv1}) into (\ref{Nn1}) leads to the sought-after expression for the resolvent
\bea
( \beta-\tfrac 12) N(\beta) \,P_V =P_V-P_V X_{+}^{-1} n X_{+}+ P_V X_{+}^{-1} n X_{-}.\label{Nn2}
\eea
We will extensively use (\ref{Nn2}). 
Throughout this section we will also make use of the following facts.
First, $n$ is a projector onto semi-positive modes, so
\bea
n|k\geq 0\rangle=|k \geq 0 \rangle, \qquad n|k<0\rangle = 0.
\eea 
And the holomorphicity of $X_\pm(x)$ in their respective domains implies that when acting on momentum states $X_-$ is a "lowering" operator and $X_+$ is a "raising" operator, i.e. the only non-zero momentum space matrix elements are
\bea
&&\langle l|X_-|m\rangle =( X_{-} )_{lm}\neq 0  \;\;\text {for }\;\; l\leq m\\
&&\langle l|X_+|m\rangle =( X_{+} )_{lm}\neq 0  \;\;\text {for }\;\; l\geq m.
\eea
This  remains valid for the inverse matrices, which retain the same upper/lower triangular form:
\bea
&&\langle l|(X_-)^{-1}|m\rangle =( X_{-}^{-1} )_{lm}\neq 0  \;\;\text {for }\;\; l\leq m\\
&&\langle l|(X_+)^{-1}|m\rangle =( X_{+}^{-1} )_{lm}\neq 0  \;\;\text {for }\;\; l\geq m.
\eea
For future reference we also give the diagonal elements
\bea
&&\langle m|X_+|m\rangle = e^{\tfrac hR(b-a)}, \qquad \braket{m|X_+^{-1}|m}=e^{-\tfrac hR(b-a)}\nn
&&\langle m|X_-|m\rangle=1,\qquad \braket{m|X_-^{-1}|m}=1.
\eea

Suppose we want to evaluate $\langle x|N(\beta)P_V|k=1\rangle$. Start with (\ref{Nn2}). Using
\bea
X_{+}^{-1} n X_{+}|k=1\rangle  =|k=1\rangle 
\eea and substituting into (\ref{Nn2}) gives 
\bea
\braket{ x|(\beta-\tfrac 12) N(\beta)P_V|k=1}=\braket{x|P_V X_{+}^{-1} n X_{-}|k=1}. 
\eea
Then, since $X_-$ is lowering, the only non-zero elements are
\bea
\braket{ x|P_V X_{+}^{-1} n X_{-}|k=1}&=&\braket{x|P_V X_{+}^{-1}|k=1}\braket{k=1| X_{-}|k=1} \nn&&+
\braket{x|P_V X_{+}^{-1}|k=0}\braket{k=0| X_{-}|k=1}. 
\eea
Matrix elements such as $\langle x|P_V X_{+}^{-1}|k\rangle $ are trivial to compute:
they equal
\bea
 \langle x|P_V X_{+}^{-1}|k\rangle=\Theta_V(x) \frac{ e^{i \tfrac kR x}}{\sqrt{2 \pi R}} X_{+}^{-1} (x).
\eea
 Since $\braket{k=1| X_{-}|k=1}=1$, the only thing left to compute is $\braket{k=0| X_{-}|k=1}$. This can be done by Cauchy's theorem and picking up the pole at infinity
\bea
\braket{k=0| X_{-}|k=1}=ih\bigg(e^{\tfrac iR b}-e^{\tfrac iR a}\bigg),
\eea
leading to
\bea
\braket{ x|N(\beta)P_V |k=1}&=&\frac{1}{\beta-\tfrac 12}\Theta_V(x) X_{+}^{-1} (x)
\frac{1}{\sqrt{2\pi R}}\bigg[e^{\tfrac iR x}+ih(e^{\tfrac iR b}-e^{\tfrac iR a})\bigg].\nn\label{Nx1}
\eea

Now let's repeat the procedure for $\braket{x|N(\beta)P_V |k\text{=-}1}$. Start again with (\ref{Nn2}). Use that $n X_{-}| k\text{=-}1\rangle =0.$
Then we need to compute
\bea
\langle x| (\beta-\tfrac 12) N(\beta)P_V |k=\text{-}1\rangle = \langle x| P_V-P_V X_{+}^{-1} n X_{+} |k=\text{-}1\rangle.\label{Nxm1}
\eea
Resist the urge to evaluate each of the two terms separately. 
Instead notice that by combining the two terms we get the complement of the projector $n$: 
\bea
n_C=1-n=\sum_{k>0} | \text{-}k\rangle \langle \text{-}k|,
\eea
and so, we have the alternative expression for (\ref{Nxm1})
\bea
\langle x| (\beta-\tfrac 12) N(\beta) P_V |k\text{=-}1\rangle = \langle x| P_V X_{+}^{-1} \,n_c\,  X_+ |k\text{=-}1\rangle.
\eea
Next, apply same reasoning as before. Since $X_+$ is raising, the only non-zero element is
\bea
\langle x| P_V X_{+}^{-1} \,n_c \,X_+ |k\text{=-}1\rangle&=& \langle x| P_V X_{+}^{-1}|k\text{=-}1\rangle \langle k\text{=-}1|X_+|k\text{=-}1\rangle \nn
&=& \Theta_V(x) X_+^{-1}(x) \frac{e^{-i\tfrac xR}}{\sqrt{2 \pi R}} e^{\tfrac hR (b-a) }.
\eea
Putting everything together,
\bea
\langle x|N(\beta) P_V|k\text{=-}1\rangle = \frac{1}{\beta-\tfrac 12}\Theta_V(x) X_+^{-1}(x) \frac{e^{-i\tfrac xR}}{\sqrt{2 \pi R}} e^{\tfrac hR(b-a) }. \label{m1m1}
\eea
Similar steps yield for example the matrix element
\bea \!\!\!\!\!\!\!
\braket{x|N(\beta)P_V|k=\text{-2}}&=&\frac{1}{\beta-\tfrac 12}\bigg(\braket{x|P_V X_+^{-1}|k=\text{-2} }
\braket{k={\text-2}|X_+|k={\text-2}}\nn
&& + \braket{x|P_V X_+^{-1}|k=\text{-1}}\braket{k=\text{-1}| X_+|k=\text{-2}}\bigg)\nn
&&
=\Theta_V(x) \frac{1}{\beta-\tfrac 12}  \frac{1}{\sqrt{2\pi R}}e^{-\tfrac {i}{R}x} e^{\tfrac hR (b-a)} X_+^{-1}(x)\bigg[ e^{-\tfrac iR x}+ih
\bigg( e^{-\tfrac iR b}- e^{-\tfrac iR a}\bigg)\bigg].\nn
\eea

As a check of the matrix elements $\braket{k=\pm 1|P_V N(\beta)|x}$ given in (\ref{n1c}) and (\ref{n-1c}), which were derived from the hermiticity of the resolvent and using (\ref{zpcc}), we offer here an alternative algebraic derivation.

To begin, from (\ref{Nn2}) by eliminating the first $P_V$ factor in terms of $X_+, X_-$ operators, we can write the resolvent as follows :
\bea
(\beta-\tfrac 12) P_V N(\beta) &=& P_V-X_-^{-1}n X_- P_V+X_+^{-1}n X_- P_V\nn
&=&X_-^{-1} n_c X_- P_V+X_+^{-1} n X_- P_V.
\eea 
This is useful then to evaluate the matrix element 
\bea
(\beta-\tfrac 12)\braket{k=\text{-1}|P_V N(\beta)|x}&=&\braket{k=\text{-1}|X_-^{-1} n_c X_- P_V|x}+\braket{k=\text{-1}|X_+^{-1} n X_- P_V|x}\nn
&=&\braket{k=\text{-1}|X_-^{-1}|k=\text{-1}}\braket{k=\text{-1}|X_-P_V|x}
\nn&=&\Theta_V(x)\frac{e^{i\tfrac xR}}{\sqrt{2\pi R}}e^{-2\pi h}z_+^{ih}(x)\nn
\eea
where we used $\langle k=\text{-1}|X_+^{-1} n =0$, and which agrees indeed with (\ref{n-1c}).

\bigskip 

Moving on to $\braket {k={\pm}1|P_V N(\beta)P_V|k={\pm}1}$ matrix elements, we have
\bea
N_{-1,-1}&\equiv&\langle k\text {=-}1|P_V N(\beta)P_V |k\text{=-}1\rangle =
\frac{1}{\beta-\tfrac 12} e^{\tfrac hR(b-a)} \braket{k=\text{-1}|P_V X_+^{-1}|k=\text{-}1}\nn
&=& e^{\tfrac hR (b-a)} -1,
\eea
where we used (\ref{m1m1}) and 
\bea
\frac{1}{\beta-\tfrac 12} \braket{k|P_V X_+^{-1}| k}=
\braket{k|(X_-^{-1} X_+-1)X_+^{-1}|k}=(1-e^{\tfrac hR (a-b)}).
\eea

Similarly,
\bea
N_{1,-1}&\equiv& \langle k=1| P_V N(\beta)P_V| k=\text {-1}\rangle = \frac{1}{\beta-\tfrac 12}
e^{\tfrac hR(b-a)} \braket{k=1|P_V X_+^{-1}|k=\text {-1}}\nn
&=&e^{\tfrac hR(b-a)} \braket{k=1|(X_-^{-1} X_+-1) X_+^{-1}|k=\text{-}1}
\eea
which requires us to evaluate $\braket{k=1|X_\pm^{-1}|k\text{=-}1}$. To this end, use Cauchy's theorem with an integration contour just inside/outside the unit circle:
\bea
&&\braket{k=1|X_+^{-1}|k\text{=-}1}=-\tfrac 12 h\bigg[(h+i)e^{-i\tfrac 2R b}+(h-i)e^{-i\tfrac 2R a}-2h e^{-\tfrac iR(a+b)}\bigg] e^{\tfrac hR (a-b)} ,\nn
&&\braket{k=1|X_-^{-1}|k\text{=-}1}=0.
\eea
This leads to
\bea
N_{1,-1} = \tfrac 12 h\bigg[(h+i)e^{-i\tfrac 2R b}+(h-i)e^{-i\tfrac 2R a}-2h e^{-\tfrac iR(a+b)}\bigg].
\eea

For the last matrix element 
\bea 
N_{1,1}\equiv \braket{k=1|P_V N(\beta)P_V |k=1}
\eea 
we use (\ref{Nx1}) to write 
\bea
N_{1,1}&=&\frac{1}{\beta-\tfrac 12}\bigg[
\braket{k=1|P_V X_+^{-1}|k=1} +ih (e^{\tfrac iR b}-e^{\tfrac iR a})
\braket{k=1|P_V X_+^{-1}|k=0}\bigg]\nn
&=&\braket{k=1|X_-^{-1}-X_+^{-1}|k=1}+ih(e^{\tfrac iR b}-e^{\tfrac iR a})
\braket{k=1|X_-^{-1}-X_+^{-1}|k=0}\nn
&=&1-e^{\tfrac hR(a-b)}-h^2\bigg|e^{\tfrac iR b}-e^{\tfrac iR a}\bigg|^2 e^{\tfrac hR(a-b)}.
\eea

From completeness, we give the matrix elements of the resolvent used in \cite{Klich:2015ina}
\bea 
&&N_{0,0}\equiv \braket{k=0|P_V N(\beta)P_V|k=0}=1-e^{-\tfrac hR (b-a)},
\nn
&&\braket{k=0|P_V N(\beta)|x}=\Theta_V(x) \frac{e^{-2\pi h}}{\beta-\tfrac 12} \frac{e^{-\tfrac hR(b-a)}}{\sqrt{2\pi R}} X_+(x),\nn
&&\braket{x|N(\beta)|k=0} = \Theta_V(x) \frac{1}{\beta-\tfrac 12} \frac{1}{\sqrt{2\pi R}}X_+^{-1}(x).
\eea 

Lastly, we will also need
\bea 
N_{-1,0}\equiv\braket{k={\text-1}|N(\beta)|k=0}=\braket{k={\text-1}|X_-^{-1}|k=0}=ih \bigg(e^{\tfrac iR b}-e^{\tfrac iR a}\bigg).\nn
\eea

%%%%%%%%%%%%%%%%%%%%%%%%%%%%%%%%%%%%%%%%%%%%%%%%%%%%%%%%%%%
\subsection{Multiple interval entangling region}\label{Apb}
%%%%%%%%%%%%%%%%%%%%%%%%%%%%%%%%%%%%%%%%%%%%%%%%%%%%%%%%%%%%%

Consider next a multiple interval entangling region: $V=\bigcup_{j=1}^N (a_j,b_j)$. This is a straightforward extension of previous section results. Starting with
the expression of $X_\pm$ on $N$ intervals (\ref{xpmninterv}),
we have to modify in our previous evaluations the matrix elements of the corresponding ''raising/lowering'' operators. We have
\bea
&&\braket{m|X_-|m}=1, \qquad \braket{m|X_-^{-1}|m}=1,\nn
&&\braket{m|X_+|m}=e^{\tfrac hR\sum_j(b_j-a_j)}, \qquad 
\braket{m|X_+^{-1}|m}=e^{-\tfrac hR\sum_j(b_j-a_j)},
\eea
and
\bea
&& \braket{k=0| X_{-}|k=1}=ih\bigg(\sum_j e^{\tfrac iR b_j}-\sum_j e^{\tfrac iR a_j}\bigg),
\nn
&& \braket{k=1|X_+^{-1}|k\text{=-}1}=
-\tfrac 12 h\bigg[h\bigg(\sum_j e^{-\tfrac iR b_j}-\sum_j e^{-\tfrac iR a_j}\bigg)^2\nn
&&\;\;\;\;\;\;\;\;\;\;\;\;\;\;\;\;\;\;\;\;\;\;\;\;\;\;\;+i\bigg(\sum_j e^{-2\tfrac iR b_j} -\sum_j e^{-2\tfrac iR a_j}\bigg) \bigg] e^{\tfrac hR \sum_j (a_j-b_j)}
\eea
In a similar fashion the matrix elements of $P_V X_\pm$ get modified to
\bea
\frac{1}{\beta-\tfrac 12} \braket{k|P_V X_+^{-1}| k}=
\braket{k|(X_-^{-1} X_+-1)X_+^{-1}|k}=(1-e^{\tfrac hR \sum_j(a_j-b_j)}).
\eea
This means that, for a multi-interval entangling region the matrix elements of the projected resolvent are
\bea
&&\langle x|N(\beta)P_V|k\text{=-}1\rangle = \frac{1}{\beta-\tfrac 12}\Theta_V(x) X_+^{-1}(x) \frac{e^{-i\tfrac xR}}{\sqrt{2 \pi R}} e^{\tfrac hR\sum_j(b_j-a_j) },\nn
&&\braket{ x|N(\beta)P_V|k=1}=\frac{1}{\beta-\tfrac 12}\Theta_V(x) X_{+}^{-1} (x)
\frac{1}{\sqrt{2\pi R}}\bigg[e^{\tfrac iR x}+ih\sum_j(e^{\tfrac iR b_j}-e^{\tfrac iR a_j})\bigg], \nn
\eea
and
\bea
N_{1,1}\equiv \langle k\text {=-}1|P_V N(\beta)P_V|k\text{=-}1\rangle
&=& e^{\tfrac hR \sum_j(b_j-a_j)} -1,\nn
N_{1,-1}\equiv\braket{k=1|P_V N(\beta)P_V |k=\text{-1}}&=&\tfrac 12 h\bigg[h\bigg(\sum_j e^{-\tfrac iR b_j}-\sum_j e^{-\tfrac iR a_j}\bigg)^2\nn
&&+i\bigg(\sum_j e^{-2\tfrac iR b_j} -\sum_j e^{-2\tfrac iR a_j}\bigg) \bigg]\nn
N_{1,1}\equiv \braket{k=1|P_V N(\beta)P_V |k=1}&=&1-e^{-\tfrac hR\sum_j(b_j-a_j)}\bigg[1+h^2\bigg|\sum_j e^{\tfrac iR b_j}-\sum_j e^{\tfrac iR a_j}\bigg|^2\bigg].\nn
\eea
\end{appendix}
\bibliographystyle{utphys}
\bibliography{RefsForEHpapers}

\end{document}